\documentclass[letter]{jpsj3}
\usepackage{graphicx}
\usepackage{latexsym}
\usepackage{fancybox}
\usepackage{color}
\title{Kondo Effects and Multipolar Order in the cubic Pr$Tr_2$Al$_{20}$ ($Tr=$Ti, V)}

\author{Akito Sakai and Satoru Nakatsuji\thanks{satoru@issp.u-tokyo.ac.jp}}  
\inst{Institute for Solid State Physics, University of Tokyo, Kashiwa, Chiba 277-8581, Japan}

\recdate{\today}

\abst{
Our single crystal study reveals that Pr$Tr_2$Al$_{20}$ ($Tr =$ Ti and V) provides the first examples of a cubic $\Gamma_3$ nonmagnetic ground doublet system that shows the Kondo effect including a $-\ln T$ dependent resistivity.
The $\Gamma_3$ quadrupolar moments in PrV$_2$Al$_{20}$ induce anomalous metallic behavior through hybridization with conduction electrons, such as $T^{1/2}$ dependent resistivity and susceptibility below $\sim 20$ K down to its ordering temperature $T_{\rm O} = 0.6$ K. In PrTi$_2$Al$_{20}$, however, quadrupoles are well-localized and exhibit an order at $T_{\rm O} = 2.0$ K.
Stronger Kondo coupling in PrV$_2$Al$_{20}$ than in PrTi$_2$Al$_{20}$ suppresses quadrupolar ordering, and instead promotes hybridization between the $\Gamma_3$ doublet and conduction electrons, leading to most likely the quadrupolar Kondo effect.
}

\kword{Pr$Tr_2$Al$_{20}$, quadrupolar Kondo effect, quadrupolar order, non-Kramers doublet}
\begin{document}
\maketitle
\newpage
The Kondo effect based on hybridization between a $4f$ moment and conduction electrons has provided a number of nontrivial phenomena in Kondo lattice systems, 
such as heavy Fermi liquid, Kondo insulator, unconventional superconductivity and quantum criticality. 
While the Kondo effect using a magnetic dipole moment is well established, 
it remains elusive whether a nonmagnetic analog is possible using multipole degree of freedom. 

Theoretically, a nonmagnetic Kondo effect is found possible using the quadrupole degree of freedom.
Cox has shown that when a $f^2$ ion such as U$^{4+}$ and Pr$^{3+}$ has a cubic site symmetry and a nonmagnetic ground doublet ($\Gamma_3$), 
the electric quadrupole moment of the ground state may interact with conduction electrons and produce renormalized properties characterized by a new small energy scale $T_{\rm QK}$ \cite{RefWorks:366}.
In this case, the quadrupolar interaction has more than one electron channel and this leads to overscreening of localized electron by conduction electrons. 
The ground state is thus non-Fermi-liquid with residual entropy of $1/2 R \ln 2$, 
exhibiting anomalous metallic behavior, such as logarithmically divergent specific heat $C_P$, $-T^{1/2}$ or $-\ln T$ dependent susceptibility, and $T^{1/2}$ dependent resistivity \cite{CoxPhysica,Kusunose}. 

Motivated by the theoretical proposal, experimental studies have been extensively made so far.
However, well-established cubic $4f^{2}~\Gamma_3$ lattice systems are still limited to a few compounds such as PrPb$_3$, PrPtBi, PrInAg$_2$, and PrMg$_3$
\cite{Morin1982257,RefWorks:364,RefWorks:71,RefWorks:335}.
The first two show quadrupolar ordering.
In particular, PrPb$_3$ is the unique example having a modulated quadrupolar phase, suggesting partial quenching
 of quadrupole moments by the hybridization \cite{Onimaru}. 
In contrast, PrInAg$_2$ and PrMg$_3$ do not exhibit any long-range order \cite{RefWorks:71,RefWorks:335}. 
Instead, for both systems $C_P/T$ saturates to a large value as $T \rightarrow 0$, as predicted for the single-channel Kondo impurity model. 
Thus, it is inferred that the local $\Gamma_3$ quadrupole moments are quenched by hybridization with conduction electrons.
It should be noted, however, that both are the Heusler-type compounds where a random site exchange is possible. 

Generally, if such disorder exists, it might lift the degeneracy of the non-Kramers $\Gamma_3$ doublet and induce a nonmagnetic ground state, 
but in the absence of the long-range order, it is hard to isolate the disorder effects. 
On the other hand, if the cubic $\Gamma_3$-doublet system exhibits a quadrupole long-range order, the degeneracy remains at least down to the transition temperature. 
Therefore, if the quadrupolar ordering temperature can be decreased by increasing the hybridization strength, one might reveal the evolution of the quadrupolar Kondo effect. 
As yet, there is no cubic $\Gamma_3$-doublet system that allows such a systematic study.

Here we report on Kondo effects and quadrupolar order found in the new cubic $\Gamma_3$ ground doublet systems PrTi$_2$Al$_{20}$ and PrV$_2$Al$_{20}$. 
A $-\ln T$ increase of the resistivity and large Weiss temperature of the susceptibility indicate the Kondo effect, which is found stronger in PrV$_2$Al$_{20}$ than the Ti counterpart. 
While PrTi$_2$Al$_{20}$ has well localized quadrupoles that order at $T_{\rm O} = 2.0$ K, quadrupolar moments in  PrV$_2$Al$_{20}$ induce anomalous metallic behavior such as divergent $C_P/T \sim T^{-3/2}$,   $-T^{1/2}$ dependent Van Vleck susceptibility, and $T^{1/2}$ dependent resistivity, above their ordering temperature $T_{\rm O} = 0.6$ K. This suggests that the increase in Kondo coupling by changing the $Tr$ site from Ti to V suppresses the quadrupolar ordering, and instead leads to stronger hybridization between the $\Gamma_3$ doublet and conduction electrons, and thus the quadrupolar Kondo effect.

The crystal structure of $RTr_2$Al$_{20}$ ($R$: rare earth, $Tr$: transition metal ) is cubic CeCr$_2$Al$_{20}$-type with the space group $Fd{\bar 3}m$ \cite{RefWorks:358}. 
Strong hybridization between $4f$ and conduction electrons is expected because $R$ ions are surrounded by sixteen Al ions, which is the largest coordination number according to Frank and Kasper
\cite{RefWorks:360}. Indeed, a nonmagnetic Ce$^{4+}$ state has been suggested for CeV$_2$Al$_{20}$ \cite{Moze199839}.  The symmetry of $R$ site is $T_{\rm d}$ and cubic. 

Single crystals of $RTr_2$Al$_{20}$ ($R=$Pr, La, $Tr=$Ti, V)
were grown by an Al self-flux method under vacuum, using 4N(99.99\%)-Pr, 3N-La, 3N-Ti,V and 5N-Al.
The crystal structure was verified by the X-ray powder and single crystal diffraction measurements and is found to have the lattice 
parameter $a = 14.723(7)$ \AA \ (Ti) and $14.591(2)$ \AA \ (V). We found no sign of randomness or distortion such as site exchange and deformation of the Al cage surrounding Pr ion \cite{Chan}.
The resistivity $\rho$ and the specific heat $C_P$ above 0.4 K were measured by a standard four-probe dc method and a relaxation method, respectively.
The dc magnetization $M$ was measured for 1.9 $<T<$ 350 K by a SQUID magnetometer.
The residual resistivity ratio (RRR) reaches up to 300 for PrTi$_2$Al$_{20}$ and up to 6 for PrV$_2$Al$_{20}$.
The residual resistivity was estimated using power law extrapolation of $\rho(T)$ to $T = 0$.
In the following, we will show the results obtained for the samples with RRR $\sim$ 300 for PrTi$_2$Al$_{20}$ and RRR $\sim$ 6 and 2 for PrV$_2$Al$_{20}$.

\begin{figure}[t]
\begin{center}
\includegraphics[keepaspectratio, scale=1.2]{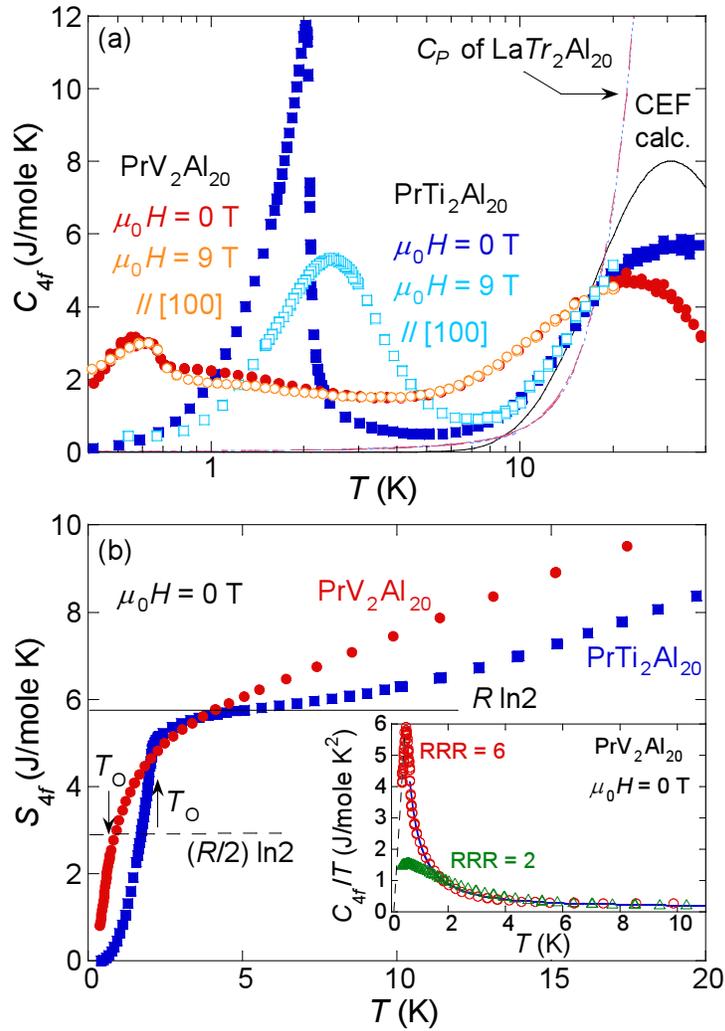}
\caption{(color online) (a) Temperature dependence of $4f$ contribution $C_{4f}$ to the specific heat $C_P$ for PrTi$_2$Al$_{20}$ (square) and PrV$_2$Al$_{20}$ (circle) under $\mu_0 H = 0$ T (solid) and 9 T (open) $ \parallel [100]$. Comparison is made with the Schottky peak calculated using the CEF scheme of PrTi$_2$Al$_{20}$ \cite{Sato_neutron}. $C_P$ for both La$Tr_2$Al$_{20}$ ($Tr = $Ti, V) are shown and collapse on top of each other. (b) Temperature dependence of $4f$ electron contribution $S_{4f}$ to the entropy. Inset: $T$ dependence of $C_{4f}/T$ for PrV$_2$Al$_{20}$ single crystals with RRR = 2 (green) and 6 (red). To estimate $S_{4f}$ at 0.4 K, a linear extrapolation (broken line) of $C_{4f}/T$ to $T = 0$  is used. Blue solid line indicates fitting to a power law.}\label{Cp}
\end{center}
\end{figure}

Figure \ref{Cp}(a) shows 4$f$ electron contribution to the specific heat, $C_{4f}$.
This was estimated by subtracting $C_P$ of the La analogs, namely, La$Tr_2$Al$_{20}$ (Fig. \ref{Cp}(a)) from $C_P$ of Pr$Tr_2$Al$_{20}$ ($Tr = $Ti, V).
A clear anomaly of $C_{\mathrm{4}f}$ is observed at $T_{\rm O} = 2.0$ K for PrTi$_2$Al$_{20}$ and $T_{\rm O} = 0.6$ K for PrV$_2$Al$_{20}$, 
indicating a 2nd-order phase transition. The weaker anomaly in PrV$_2$Al$_{20}$ suggests the stronger hybridization effect, as we will discuss later.
To estimate the entropy $S_{\mathrm{4}f}$, we integrated $C_{4f}/T$ vs. $T$ above 0.4 K.
For PrV$_2$Al$_{20}$, a linear decrease of $C_{4f}/T$ below $T_{\rm O}$ was assumed to estimate  $S_{\mathrm{4}f} \sim 0.8 $ (J/
mole K) at 0.4 K (Fig. \ref{Cp}(b) inset). $S_{\mathrm{4}f}$ for both systems shows a saturation to $\sim R\ln2$ at $T \sim 5$ K, indicating the ground doublet (Fig. \ref{Cp}(b)). 
No Schottky anomaly due to the Zeeman effect of $4f$ moments was observed in $C_{4f}$ under a field of 9 T for both systems.
$\Gamma_4$ and $\Gamma_5$ are the magnetic triplets under the cubic crystalline electric field (CEF) and have the respective magnetic moments of $0.4$ and $2 \mu_{\rm B}$.
Insensitivity of $C_{4f}$ to the magnetic field (Fig. \ref{Cp}(a)) rules out such magnetic CEF ground states and indicates the nonmagnetic ground doublet $\Gamma_{3}$.

Indeed, inelastic neutron measurements for PrTi$_2$Al$_{20}$ confirm the $\Gamma_3$ ground doublet and that the excited $\Gamma_4$, $\Gamma_5$, and $\Gamma_1$ states are separated from $\Gamma_3$ by 5.7, 9.5 and 13.6 meV, respectively \cite{Sato_neutron}. 
This scheme roughly reproduces the CEF Schottky anomaly seen as a broad peak of $C_{4f}$ at $\sim 30$ K in PrTi$_2$Al$_{20}$ (Fig. \ref{Cp}(a)). 
Most likely strong hybridization between $4f$ and conduction electrons  broadens the peak in comparison with the CEF calculation, as we will discuss. 
Likewise, PrV$_2$Al$_{20}$ has a broad peak at $\sim 20$ K, indicating that the CEF gap $\Delta$ to the first excited state is $\sim 40$ K.  
Given $T_{\rm O} \ll \Delta$ for both systems, the phase transitions must be caused by multipolar ordering of the $\Gamma_3$ doublet. 

Although the transition temperature in PrTi$_2$Al$_{20}$ is almost the same for all the crystals having various RRR between 20 and 300, the transition in PrV$_2$Al$_{20}$ is found strongly sensitive to the sample quality.
While a clear transition anomaly is found at $T_{\rm O} = 0.6$ K for the PrV$_2$Al$_{20}$ crystals with RRR $> 4$,
$C_{4f}$ of the low quality sample with RRR = 2 does not exhibit any anomaly, indicating the absence of the transition at $T > 0.4$ K (Fig. \ref{Cp}(b) inset). 
Instead, $C_{4f}/T$ gradually saturates to $\sim 1.5$ J/mol$K^2$ as $T \rightarrow 0$. 
This large $C_{4f}/T$ value should not be attributed to the formation of heavy Fermi liquid, but to the random freezing of the quadrupole moments. 

\begin{figure}[t]
\begin{center}
\includegraphics[keepaspectratio, scale=1.2]{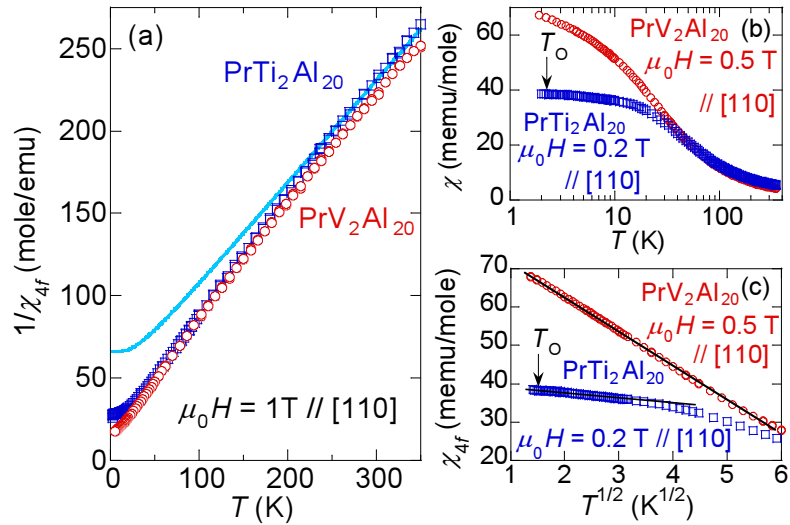}
\caption{(color online) (a) Temperature dependence of the inverse of $4f$ electron contribution $\chi_{4f}$ to the susceptibility $\chi = M/H$ of PrTi$_2$Al$_{20}$ (square) 
and PrV$_2$Al$_{20}$ (circle) under $\mu_0 H =  1$ T $\parallel [110]$. 
Definition of $\chi_{4f}$ is given in the text. The CEF $\chi^{-1}$ (solid line) for PrTi$_2$Al$_{20}$ is computed using the CEF scheme obtained by the neutron 
diffraction \cite{Sato_neutron} and an exchange field parameter $\lambda = -34.0$ mole/emu. Right panels: 
(b)$\chi$ vs.  $T$ and (c)$\chi_{4f}$ vs. $T^{1/2}$ obtained under low fields applied along $[110]$.}\label{chi}
\end{center}
\end{figure}

Figure \ref{chi}(a) shows the $T$ dependence of the inverse of the 4$f$ electron contribution to the susceptibility $\chi_{4f}$,
 defined as the difference between $M/H$ of Pr$Tr_2$Al$_{20}$ and La$Tr_2$Al$_{20}$. 
$M/H$ of La$Tr_2$Al$_{20}$ shows Pauli susceptibility type behavior with nearly constant $\chi \sim 0.4$ (Ti) and $-0.6$ (V) memu/mol as a function of $T$.
$\chi_{4f}$ is nearly field independent and isotropic under $\mu_0 H \leq 1$ T above $T_{\rm O}$.
Above 250 K, $\chi_{4f}$ can be fit to the Curie-Weiss formula, $1/\chi_{4f} = (T-\Theta_{\rm W})/C$ and yields an antiferromagnetic Weiss temperature,
 $\Theta_{\mathrm{W}}=-40$ and $-55$ K and effective moments of $\mu _{\mathrm{eff}} = 3.43$ and $3.57 \mu _\mathrm{B}$ for PrTi$_2$Al$_{20}$ and PrV$_2$Al$_{20}$, respectively. $\mu _{\mathrm{eff}}$ is consistent with the theoretical value expected for the full multiplet of Pr$^{3+}$ ($3.58 \mu _\mathrm{B}$). 
However, given the large Pr-Pr interspacing $\sim 6.7$ \AA, $\Theta_{\mathrm{W}}$ is too large for the RKKY interaction, and thus should represent the Kondo coupling scale.

$\chi_{4f}(T)$ of PrTi$_2$Al$_{20}$ is nearly constant below $T = 30$ K $\sim \Delta/2$ (Fig. \ref{chi}(b)), as expected for the nonmagnetic $\Gamma_3$ ground doublet with the CEF gap $\Delta$. 
$\chi(T)$ computed using the above CEF scheme and an exchange field parameter $\lambda = -34.0$ mole/emu well reproduces the experiment, except the slope change in $1/\chi_{4f}$ at $T \sim 230$ K (Fig. \ref{chi}(a)). The slope increase may correspond to the decrease in the effective moment size, and thus suggests a partial screening of $4f$ moments due to the Kondo effect.
Although more strongly $T$ dependent, $\chi_{4f}(T)$ of PrV$_2$Al$_{20}$ starts saturating at $T \sim 20$ K $\sim \Delta/2$, indicating the onset of the Van Vleck paramagnetism due to the $\Gamma_3$ ground doublet. 
As we will discuss, the enhancement below 20 K may well be a hybridization effect between the $\Gamma_3$ doublet and conduction electrons.
Besides, the larger slope of $1/\chi_{4f}$ of PrV$_2$Al$_{20}$ than the Ti case at $T < 100$ K suggests stronger screening effects.

\begin{figure}[t]
\begin{center}
\vspace{40pt}
\includegraphics[clip]{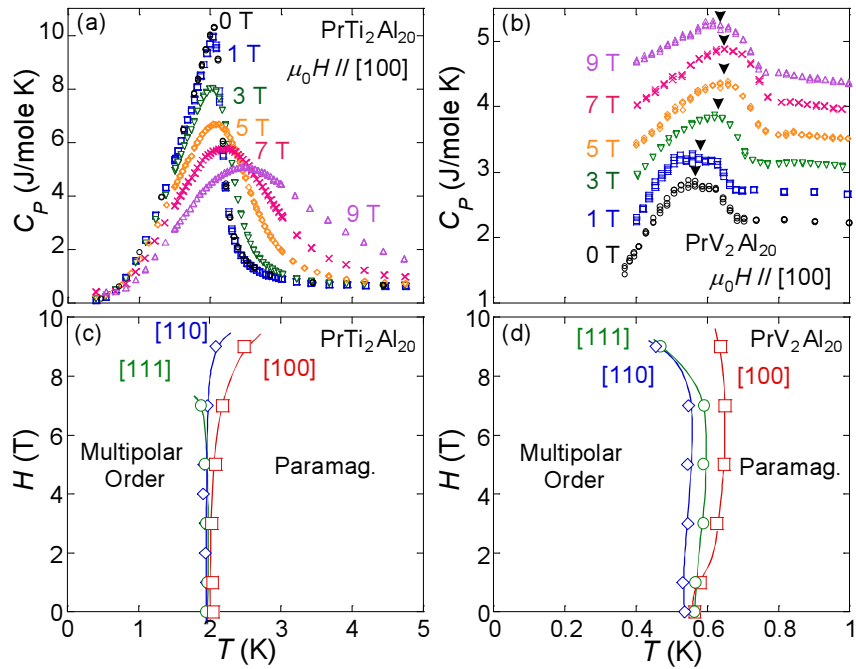}
\caption{(color online) Temperature dependence of the specific heat $C_P$ under various  fields applied along $[100]$ for (a) PrTi$_2$Al$_{20}$ and (b) PrV$_2$Al$_{20}$. 
For PrV$_2$Al$_{20}$, values of $C_P$ under fields are shifted upwards for clarity.
$H$-$T$ phase diagrams for (c) PrTi$_2$Al$_{20}$ and (d) PrV$_2$Al$_{20}$. 
$\Box$, $\Diamond$ and $\bigcirc$ indicate the peak positions of $C_P$ 
under $H \parallel [100], [110]$ and [111], respectively. Solid lines are guides to the eye.} 
\label{Cp-B}
\end{center}
\end{figure}

Now, we discuss the low $T$ phase transition, having established that the ground doublet is the cubic $\Gamma_3$.
Figures \ref{Cp-B}(a) and (b) show $C_P(T)$ under various fields applied along [100] for PrTi$_2$Al$_{20}$ and PrV$_2$Al$_{20}$, respectively. 
With increasing field, the $C_P(T)$ peak shifts to a higher temperature.
While the peak keeps nearly the same shape under field for PrV$_2$Al$_{20}$ as expected for an antiferroquadrupolar ordering, it becomes broader for PrTi$_2$Al$_{20}$, suggesting that the transition is not well-defined under field. 
This type of crossover is not allowed for an antiferro type, but for a ferroquadrupolar order as in  PrPtBi \cite{RefWorks:364}, because
only the symmetry in a ferroquadrupolar state can be the same as the one in a para state under field.
Therefore, the multipolar transitions in PrTi$_2$Al$_{20}$ and PrV$_2$Al$_{20}$ are most likely a ferro- and antiferro-quadrupolar ordering, respectively.


Further evidence for quadrupolar order comes from $H$-$T$ phase diagrams of both PrTi$_2$Al$_{20}$ and PrV$_2$Al$_{20}$ under fields along $[100]$, $[110]$ and $[111]$, as shown in Figs. \ref{Cp-B}(c) and (d).
The transition temperature $T_{\rm O}(H)$ is determined as the peak $T$ of $C_P(T)$.
In both systems, $T_{\rm O}(H)$ increases with increasing field in the low field regime, except for $H \parallel [111]$ in PrTi$_2$Al$_{20}$. 
This increase of the boundary with field is characteristic to quadrupolar order, and is the effects of dipole moments induced by magnetic field, which assist the quadrupolar ordering \cite{Shiina1997,Kuramoto}.

\begin{figure}[t]
\begin{center}
\includegraphics[keepaspectratio, scale=1.2]{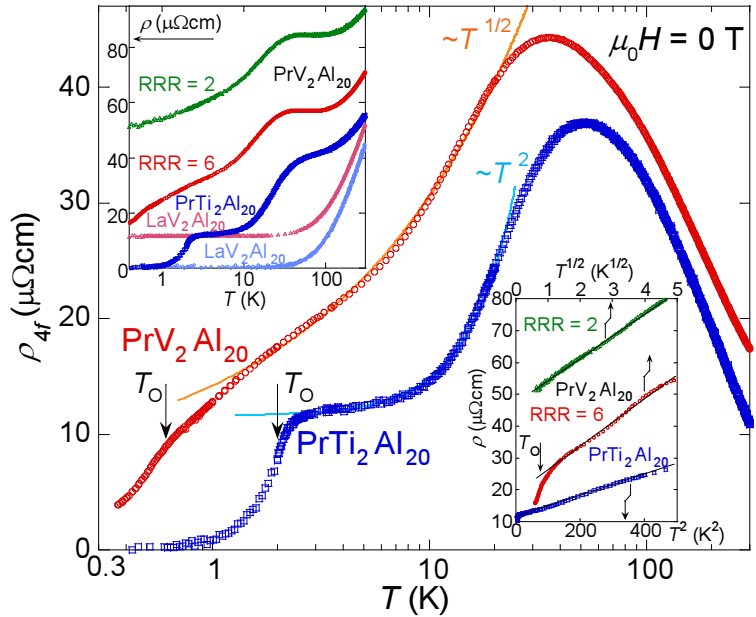}
\caption{(color online) $4f$ electron contribution $\rho_{4f}(T)$ to the resistivity $\rho(T)$ of Pr$Tr_2$Al$_{20}$ ($Tr =$ Ti and V). Arrows indicate the transitions at $T_{\rm O} = 0.6$ K (V) and 2.0 K (Ti). The solid lines indicate fits to the power law $T$ dependence, namely $\Delta \rho_{4f} \sim T^{1/2}$ (V) and $\Delta \rho_{4f} \sim T^2$ (Ti).
Left inset: $\rho(T)$ for Pr$Tr_2$Al$_{20}$ and La$Tr_2$Al$_{20}$.
$\rho_{4f}$ is defined as the difference between the two for each $Tr$. Right inset: low $T$ $\rho$ vs. $T^{1/2}$ (V) and $T^2$ (Ti). For PrV$_2$Al$_{20}$, results for single crystals with RRR = 2 (green) and 6 (red) are shown.
}\label{rho}
\end{center}
\end{figure}


Now, we consider the paramagnetic phase above $T_{\rm O}$ to discuss the Kondo effect.
Figure \ref{rho} shows the 4$f$ electron contribution to the resistivity, $\rho_{4f}(T)$, obtained after subtracting $\rho(T)$ of the La analog. 
The raw data of $\rho(T)$ are also shown in the left inset of Fig. \ref{rho}. 
Both Pr$Tr_2$Al$_{20}$ systems exhibit a resistivity drop due to the transition at $T_{\mathrm O} = 2.0 $ K (Ti) and 0.6 K (V).
Interestingly, $\rho_{4f}$ shows a $-\ln T$ dependence at high $T$ and forms a peak at $\sim 55$ K (Ti), and $\sim 40$ K (V).
This is reminiscent of the Kondo effects seen in Ce, Yb, and U-based heavy-fermion compounds \cite{HewsonBook}.
The CEF scheme alone cannot produce the $-\ln T$ dependence but only the $T$ independent $\rho$ at $T > \Delta$. 
Because $\rho_{4f}(T)$ peaks around $T$ corresponding to the CEF $\Delta \sim 60$ K (Ti) and $\sim 40$ K (V), the $-\ln T$ dependence 
may well come from the Kondo effect using the excited magnetic $\Gamma_4$ and/or $\Gamma_5$ states.
Such Kondo effect has been found in a few Pr materials i.e. PrSn$_3$, PrFe$_4$P$_{12}$, and Pr$_2$Ir$_2$O$_7$ \cite{PrSn3_1,PrSn3_2,PrFe4P12,PrIr2O7},  
and the current systems are the first examples of a cubic $\Gamma_3$ nonmagnetic ground doublet system that shows the Kondo effect. 
Thus, we expect hybridization effects between the $\Gamma_3$ doublet and conduction electrons.

Indeed, PrV$_2$Al$_{20}$ exhibits strong hybridization effects below $\sim 20$ K $< \Delta$, where the dominant character of $4f$ moments is quadrupolar.
Most prominent feature is the divergent increase in $C_{4f}/T$, which shows $\sim 1/T^{\alpha}$ dependence below 10 K with $\alpha \sim 1.5$, and reaches $\sim 5$ J/mole-Pr K$^2$ at $\sim T_{\mathrm O}$ (Fig. \ref{Cp}(b) inset).
The $T$ range is too wide to ascribe this enhancement to critical fluctuations associated with the transition at $T_{\rm O} = 0.6$ K. 
In addition, $\chi_{4f}(T)$ of PrV$_2$Al$_{20}$ shows pronounced $-\sqrt T$ dependence over a decade of $T$ between 2 K and $ \sim 30$ K (Fig. \ref{chi} (c)). 
This cannot be understood in terms of a CEF effect, which only produces a $T$ independent Van Vleck $\chi$.
Actually, this type of anomalous increase in the Van Vleck paramagnetic regime has been seen in other cubic Pr compounds such as PrMg$_3$ \cite{Morie}, and is ascribed to the hybridization effect between $4f \Gamma_3$ and conduction electrons, 
as the theory for the quadrupolar Kondo effect predicts $-\sqrt T$ dependence  \cite{CoxPhysica}.
Furthermore, $\rho(T)$ shows $T^{1/2}$ dependence between $\sim 1.5$ K and 20 K (Fig. \ref{rho} and its right inset),
again as predicted for the quadrupolar Kondo effect \cite{CoxPhysica}. 
The $T^{1/2}$ dependence in $\chi(T)$ and $\rho(T)$ is found in all the samples with various sample quality (e.g. see Fig. \ref{rho} inset).


In contrast, quadrupoles in PrTi$_2$Al$_{20}$ are more localized than in PrV$_2$Al$_{20}$.
This is inferred from the small amount of entropy release $\sim 10$ \% of $R\ln 2$ above $T_{\rm O}$. 
A weak $T$ dependence in $\rho(T)$ below 10 K down to $T_{\rm O}$ also suggests local fluctuations of localized quadrupoles. 
Although weak, the system still exhibits hybridization effects. 

$\chi_{4f}$ shows $-\sqrt T$ dependence at lower $T~(<$ 9 K)  than in PrV$_2$Al$_{20}$ (Fig. \ref{chi} (c)).
On the other hand, $\rho(T)$ fits to a $T^2$ law at $T < 20$ K (Fig. \ref{rho} and its right inset), 
which suggests mass enhancement of conduction electrons through the hybridization.



The above difference in the hybridization effects should come from the following reasons.
First, the replacement of Ti by V compresses the unit cell volume by $\sim 3$ \%, and thus plays the role of chemical pressure. 
Secondly, V  has an additional $3d$ electron in comparison with Ti, which should contribute to the conduction band.
The stronger hybridization effects seen in PrV$_2$Al$_{20}$ than PrTi$_2$Al$_{20}$ at $T < 20$ K are consistent with 
the high $T$ properties such as the  higher 
$\Theta_{\rm W}$ and 
the stronger anomaly due to the Kondo effect in $\rho_{4f}$ in PrV$_2$Al$_{20}$ (Fig. \ref{rho} and its left inset), both of which point to the larger Kondo coupling in PrV$_2$Al$_{20}$ than the Ti counterpart. 

Because the $\Gamma_3$ doublet degeneracy is not lifted down to $T_{\mathrm O}$, 
the Kondo effect using the quadrupole degree of freedom such as quadrupolar Kondo effects may well be the origin of the anomalous metallic behavior seen in PrV$_2$Al$_{20}$. 
In fact, the $\sqrt T$ dependence seen in both $\chi_{4f}(T)$ and $\rho_{4f}(T)$ is consistent with the theoretical prediction \cite{CoxPhysica}. 
In addition, while the observed $T^{-3/2}$ dependence of $C_{4f}/T$ is more divergent than the $-\ln T$ divergence predicted by the theory, 
the entropy release below $T_{\rm O}$ is close to the theoretical amount of the residual entropy $\sim 1/2 R \ln 2$ (Fig. \ref{Cp} (b)).
This suggests that the quadrupolar Kondo effects are dominant at $T > T_{\rm O}$, suppressing the instability of quadrupolar order by strongly screening the quadrupole moments \cite{RefWorks:366}. The fact that the $\sqrt T$ dependence of both $\chi_{4f}(T)$ and $\rho_{4f}(T)$ is seen irrespectively of RRR points to the single ion character of the quadrupolar Kondo effect. 
On the other hand, the anomalous metallic behavior seen up to relatively high temperature of $\sim 20$ K in PrV$_2$Al$_{20}$ suggests the Kondo scale $T_{\rm QK} \sim 100$ K, according to the theory for the single ion quadrupolar Kondo effect \cite{CoxPhysica}. This is apparently too high for the theoretical model, and calls for the theoretical extension to include the lattice and CEF effects.

Finally, 
the fact that the increase in the Kondo coupling in turn decreases the ordering temperature and leads to anomalous metallic state suggests the proximity to a quantum critical point separating a quadrupolar ordered state 
and a disordered ground state.
As far as we know, no such possibility has been discussed either experimentally or theoretically.
The study of the Pr$Tr_2$Al$_{20}$ family might allow us to find such a novel quantum critical point and may unveil further exotic phenomena.

\begin{acknowledgment}
We thank Julia Y. Chan, M. Kangas, D. C. Schmitt, T. Sakakibara, T. Tomita, E.T.C. O'Farrell, T. J. Sato, Y. Nakanishi, K. Hattori and K. Ueda for useful discussions.
This work is partially supported by Grants-in-Aid (No.21684019) from 
JSPS, by Grants-in-Aids for Scientific Research on Innovative Areas ``Heavy Electrons" of MEXT, Japan. 
\end{acknowledgment}
\bibliographystyle{jpsj}

\end{document}